\documentstyle[12pt]{article}  

\begin{document}

\textwidth=6.0 in
\textheight=8.5 in
\linewidth=6.0 in
\voffset=-20mm
\hoffset=-10mm
\vsize=8.5 in 
\hsize=6.0 in

  
\begin{flushright}
{Alberta Thy 17-99}
\end{flushright}
 
\begin{center}  
  
{\Large\bf Two-dimensional effective action\\ 
for matter fields coupled to the dilaton}  
\end{center}  
\vspace{7mm}  
  
\baselineskip= 6mm  
\begin{center}  
{\bf Yu. V. Gusev}\\  
\vspace{3mm}  
{\em Theoretical Physics Institute,   
University of Alberta,\\  
Edmonton, Alberta, Canada T6G 2J1\\  
e-mail: ygusev@phys.ualberta.ca}   
\vspace{10mm}  
     
{\bf  A. I.  Zelnikov}\\  
\vspace{3mm}  
{\em Theoretical Physics Institute,   
University of Alberta,\\  
Edmonton, Alberta, Canada T6G 2J1\\  
e-mail: zelnikov@phys.ualberta.ca}\\  
and\\  
{\em P.N. Lebedev Physics Institute,\\  
Leninskii prospect 53, Moscow 117 924 Russia}  
\end{center}

\centerline{\bf Abstract}    
             
\noindent  
{\small	  
We revise the calculation of the one-loop  
effective action for scalar and spinor fields coupled to the dilaton  
in two dimensions. Applying the method of covariant  
perturbation theory for the heat kernel   
we derive the effective action in an explicitly  
covariant form that produces both the conformally invariant   
and the  conformally anomalous terms.   
For scalar fields the conformally invariant part of the action is   
nonlocal. The obtained effective action is proved to be   
infrared finite. We also compute the one-loop effective action  
for scalar fields at finite temperature.  
}

  
\thispagestyle{empty}  
\pagebreak  


\section{Introduction}  
  
The effective action and conformal anomaly of quantum fields  
coupled to the dilaton in two dimensions have been the subject of
a number of recent papers. The main motivation   
for the study of quantum field models in 2D dilaton-gravity backgrounds 
comes from the fact that such models naturally arise after the spherical  
(or dimensional) reduction from higher-dimensional field theories 
and gravity. For a  description of the spherical reduction procedure  
leading to dilaton gravity in two dimensions, 
we refer to \cite{RussoSusskThorl-PLB92,KummLieblVass-1997}.

Two dimensional models seem to be easy   
to quantize, and in some cases they admit exact solutions
at classical and quantum levels. 
Black hole physics is one of the most interesting 
applications of such models
\cite{RussoSusskThorl-PLB92,CGHS-PRD92}.
Seemingly, the two-dimensional results could 
provide information about higher-dimensional quantum physics. 
The question of the effective action for 2D dilaton gravity 
and its relation to Hawking radiation is addressed in many papers   
(\cite{MukhWipfZel-PLB94,KummVass-hepth9811,%
BalbFabbri-PRD99,LombMazzRusso-PRD99}, to mention a few). 
For the history and contemporary state 
of this problem see a recent review by Kummer and Vassilevich 
\cite{KummVass-grqc9907}.  
However, the applicability of these two-dimensional considerations
to the Hawking effect in four dimensions
is hampered by serious problems   
\cite{BalbFabbri-PRD99,BalbFabbri-grqc99,NojiriOdin-PLB92}  
(some of these problems are related to 
the Dimensional-Reduction Anomalies and may be resolved   
via their thorough analysis \cite{FrolSuttZeln-hepth99}).

We study  here only two-dimensional models,
so, we are not concerned with problems 
related to the dimensional reduction
or higher-dimensional quantum physics. 
Specifically, in this paper we focus on 
the one-loop effective action   
for quantum matter fields interacting with the background dilaton 
and gravity in two dimensions, and related infrared problems.  
Surprisingly, there is no consensus in the literature even about this   
relatively simple problem. We derive the one-loop effective action,
which, in our opinion, corrects and supplements other known results
on this subject.

  
\section{Heat kernel for scalar fields   
coupled to the dilaton in two dimensions}  
\label{computation}  
  
  
Let us begin with the classical action  
for the scalar matter field $\eta$ coupled to the background   
metric $g_{\mu\nu}$ and the background dilaton field $\phi$,  
\begin{equation}   
S = - \frac12 \int {\mathrm d}^2 x \, g^{1/2}  
    {\mathrm e}^{-2 \phi} \nabla^{\mu} \eta  
    \nabla_{\mu} \eta.                 \label{model}  
\end{equation}  
We do not specify here the function $\phi$, which can be an arbitrary
smooth function.  
Following the procedure of   
\cite{MukhWipfZel-PLB94,Dowker-CQG98}   
we redefine field variables and rewrite the action (\ref{model})   
in terms of new scalar fields   
$\tilde{\eta} = {\mathrm e}^{-\phi} \eta$.
Then the action takes the form  
\begin{equation}   
S = - \frac12 \int {\mathrm d}^2 x \, g^{1/2}   
    \Big\{  
    \nabla^{\mu} \tilde{\eta} \nabla_{\mu} \tilde{\eta}  
	 - \tilde{\eta}^2 [ \Box \phi   
	- ({\nabla}_{\mu} \phi) ({\nabla}^{\mu} \phi)]  
    \Big\}. \label{newmodel}  
\end{equation}
The one-loop effective action for this model is  
defined as  
\begin{equation} 	  
W = \frac12\, {\mathrm Tr}\, \ln\, F (\nabla), \label{trln}  
\end{equation}
where the differential  
operator corresponding to the action (\ref{newmodel}) reads  
\begin{equation}   
    F(\nabla)= 
        \Box  + \Box \phi - 
        ({\nabla}_{\mu} \phi) 
        ({\nabla}^{\mu} \phi).  \label{operator}  
\end{equation}

The widely accepted technique to compute 
the effective action is to use the trace anomaly 
of the energy-momentum tensor (Weyl anomaly),  
$T=2 g^{\mu\nu} (\delta W/ \delta g^{\mu\nu})$.
Combined with the proper boundary conditions it provides
enough information to derive unambiguously the one-loop effective
action in the absence of the dilaton, 
the Polyakov action \cite{Polyakov-PLB81}. 
A similar method was applied 
to the system of quantum scalar fields coupled to the dilaton
\cite{MukhWipfZel-PLB94,Dowker-CQG98,ChibaSiino-MPLA97,%
BoussoHawk-PRD97,NojiriOdin-MPL97}.
The operator (\ref{operator}) describes a 2D conformal model,
and, in this case,  $W$ is also restored 
by the  integration of  the conformal anomaly.
Such an effective action is known as the anomaly-induced action.
Because this action is completely defined by T, 
the main subject of calculations and controversies
in the existing literature was the computation of the anomaly itself.   
Unfortunately, unlike the Polyakov action, 
the anomaly-induced action is incomplete 
because it may contain conformally invariant terms that
can not be fixed by knowledge of the anomaly alone. 
These missing terms are important, for they lead to 
a non-zero (though traceless) energy-momentum tensor.
This ambiguity is an artifact of the method, and 
its origin is obvious. The integral of the Weyl anomaly
is, in fact, the difference between the effective actions in a physical
spacetime and in a reference one \cite{Dowker-CQG94}. 
Without a dilaton the reference spacetime is implicitly assumed 
to be flat with the same topology as the physical 2D manifold. 
The presence of the dilaton leads to 
a nontrivial conformally invariant effective action 
in the reference spacetime. 
As we will show explicitly, this action is generically nonlocal; 
hence, it can contribute to the Hawking 
radiation from the 2D dilaton black holes.  
  
In order to obtain the complete effective action we 
use a method, which is different from the one we just described. 
Our approach to this problem has two important features: 
1) it is manifestly covariant  throughout all calculations; 
2) it does not make use of the trace anomaly, 
thus, both anomaly-producing and conformally invariant  
terms come from the same calculation. 

We begin with the heat kernel for the operator (\ref{operator})  
and express the one-loop effective action  
as an integral  over the proper time $s$,  
\cite{DeWitt-book65,BarVilk-PRep85},    
\begin{equation}  
	W = - \frac12 \int^\infty_0
	\frac{{\mathrm d} s }{s} {\mathrm Tr} K(s), \label{efac} 
\end{equation}  
In coordinate representation ${\mathrm{Tr}}$ 
denotes the functional trace,  
$	{\mathrm Tr} \, K(s)=\int  {\mathrm d}^D x \,   
{\mathrm tr} \hat{K}(s|x,x)$  
in arbitrary dimensions $D$,   
where  $~{\mathrm tr}~$ denotes the matrix trace over any  
internal degrees of freedom that may be present in a field theory.    
The heat kernel $\hat{K}(s)$ is defined as a solution of the problem  
\begin{eqnarray}  
	\frac{\mathrm d}{{\mathrm d} s} \hat{K} (s|x,y) =  
    \hat{F}(\nabla^x) \hat{K}(s|x,y), \hskip 2cm   
    \hat{K}(0|x,y)=\hat{1} \delta(x-y). \label{heateq}  
\end{eqnarray}    
  
For the computation of the heat kernel we employ   
the covariant perturbation theory of Barvinsky and Vilkovisky  
\cite{CPT1,CPT2,CPT3,CPT4}.  
As a basis for our calculations we use  
a general expression for the trace of the heat kernel   
in arbitrary spacetime dimensions  
obtained in Refs.~\cite{CPT4,BGVZ-JMP94-asymp,BGVZ-JMP94-basis}  
up to the third order in curvatures,  
\begin{eqnarray}  
	{\mathrm Tr} K(s) &=&   
    \frac1{(4\pi s)^{D/2}}\int\! {\mathrm{d}}^{D}x\, g^{1/2}\,   
\Big\{  
    1+ s \hat{P}  
	+s^2 \sum_{i=1}^5  
	 f_{i}(-s \Box_2) \Re_1 \Re_2 (i)  
\nonumber\\[2mm]&& \mbox{}      
     + s^3 \sum_{i=1}^{29}  
    F_i (- s\Box_1, -s \Box_2, -s \Box_3) 	  
    \Re_1 \Re_2 \Re_3 (i)  
	 + {\mathrm O}[\Re^4]   
\Big\},  				\label{TrK}  
	\end{eqnarray}  
for the generic field operator  
\begin{equation}   
    \hat{F} (\nabla)=   
    \hat{1} \Box -   \frac{\hat{1}}{6} R(x) + \hat{P}(x).  
    				\label{op}  
\end{equation}  
Here $R$ is the Ricci scalar, and $\hat{P}$ 
is an arbitrary potential term,  
which depends on the background fields and curvatures.  
In (\ref{TrK}) we introduced the collective notation for background   
field strengths (``curvatures''),   
$\Re= (R_{\mu\nu}, \hat{\cal R}_{\mu\nu}, \hat{P})$,  
which includes the commutator curvature,  
\begin{equation}  
    [ \nabla_{\mu}, \nabla_{\nu}]  \eta  =  
     \hat{\cal R}_{\mu\nu} \eta.  \label{commutator}  
\end{equation}

The form factors $f_i$ and $F_i$ in Eq.~(\ref{TrK})   
are analytic functions of the dimensionless argument   
	$s\Box$  
that act on tensor invariants constructed from  
the field strengths.  
It is assumed that the operator arguments $\Box_i$  
in the form factors   
are acting on the curvatures at the corresponding   
spacetime points, $\Re_i = \Re (x_i)$,  
and after that all spacetime points are made coincident,  
$x_1=x_2=x_3=x$.

For straightforward applications of this result,  
the differential operator for a field model   
should be of the form (\ref{op}).   
It was already shown \cite{CPT4,BGVZ-JMP94-asymp} 
that the heat kernel (\ref{TrK})
correctly reproduces the Polyakov action,
where the operator is just $ F(\nabla) = \Box $.
The operator (\ref{operator}) also belongs to the class   
of models  (\ref{op}) with the following specifications,  
\begin{equation}  
    {\mathrm tr} \hat{1}=1,  
    \hspace{7mm} \hat{P}= \Big(\frac16 R  + \Box \phi   
    - (\nabla_\mu \phi)   
	(\nabla^{\mu} \phi) \Big) \hat{1}. \label{potential}  
\end{equation}   
Furthermore, the basis of 29 tensor structures  
in the third order \cite{CPT4,BGVZ-JMP94-basis} can be considerably
reduced using the identities   
\begin{equation}  
    \hat{\cal R}_{\mu\nu} =0,   
    \hspace{7mm}   
    R_{\mu\nu}=\frac12 g_{\mu\nu} R.  
\end{equation}  
It is useful  to express the heat kernel (\ref{TrK})  
in terms of two background field objects,  
the Ricci scalar $R$ and the dilaton field $\phi$,  
instead of $R$ and $P$.  Integrating by parts and 
discarding total derivatives   
we represent the first local term of (\ref{TrK})  
in the form  
\begin{equation}  
  \int\! {\mathrm d}^2 x\, g^{1/2}\,  P(x) =  
\int\! {\mathrm d}^2 x\, g^{1/2}\,    
	\phi \Box \phi. \label{P}  
\end{equation}  
The expression for the trace of the heat kernel  
for operator (\ref{operator}) reads  
\begin{eqnarray}  
    {\mathrm Tr} K(s) &=&  
    \frac1{4\pi s}\int\! {\mathrm d}^2 x\, g^{1/2}\,    
\left\{  
    1  
    + s  \phi \Box \phi   
    + s^2   
\left[  
    \frac12  f(-s {\Box_2}) (\Box {\phi}_1)(\Box {\phi}_2)   
\right.  \right.  
\nonumber\\&&\mbox{}  
   +   
\left(   
    \frac1{32}f(-s\Box_2)-\frac18  
    \Big(\frac{f(-s\Box_2)-1}{s\Box_2}\Big)  
\right.   
\nonumber\\&&\mbox{}  
\left.  
    +\frac38  
  \Big(\frac{f(-s\Box_2)-1-\frac16s\Box_2}{(s\Box_2)^2}\Big)  
    \right) R_1 R_2  
\nonumber\\&&\mbox{}  
    + \Big( \frac14  f(-s {\Box_2})    
    -  \frac12  \frac{f(-s {\Box}_2) -1}{s \Box_2}    
    \Big)  
    (\Box {\phi}_1 - (\nabla {\phi}_1)^2  ) R_2   
\Big]   
\nonumber\\&&\mbox{}  
    + s^3 \Big[   
    M_1 (- s\Box_1, -s \Box_2, -s \Box_3) 	R_1 R_2 R_3   
\nonumber\\&&\mbox{}  
    + M_2 (- s\Box_1, -s \Box_2, -s \Box_3)    R_1 R_2 (\Box \phi_3 )  
\nonumber\\&&\mbox{}  
    + M_3 (- s\Box_1, -s \Box_2, -s \Box_3) 
	R_1 (\Box {\phi_2}) (\Box \phi_3)   
\nonumber\\&&\mbox{}  
    + M_4 (- s\Box_1, -s \Box_2, -s \Box_3)  
    (\Box {\phi_1}) \left. (\Box {\phi_2}) \vphantom{\frac12}   
    (\Box \phi_3)\Big]  
    + {\mathrm O}[\Re^4]  \right\}. \label{HK}  
\end{eqnarray}  
In any expression, which depends on $\phi$ and  $R$,
like Eq.~(\ref{HK}), we assume $\Re= (\phi, R)$.
All second-order form factors 
are expressed via the basic one  
\begin{equation}   
	f(-s\Box)=\int_0^1\!\! {\mathrm d}\alpha\:    
	{\mathrm{e}}^{\alpha(1-\alpha)s\Box}.   \label{basicf}  
\end{equation}   
The third-order form factors $M_i$ for $i=1 \ldots 4$  
are functions of the dimensionless arguments   
$\xi_k = - s\Box_k $, $k=1 \ldots 3$  
and are listed in Appendix A.   
They are formed with the basic form factors    
(\ref{basicf}) and   
\begin{eqnarray}  
F (-s\Box_1,-s\Box_2,-s\Box_3)   
&=&  
   \int_{\alpha\geq 0} {\mathrm d}\alpha_1   
\,{\mathrm d}\alpha_2 \,{\mathrm d}\alpha_3\,  
\delta(1-\alpha_1-\alpha_2-\alpha_3)
\nonumber\\&& \mbox{}  
 \times  \exp \Big( s ( \alpha_2 \alpha_3 \Box_1 +   
\alpha_1 \alpha_3 \Box_2 + \alpha_1 \alpha_2 \Box_3 )  
\Big).  
\end{eqnarray}         
The form factors in (\ref{TrK}) and (\ref{HK})  
are analytical functions of the proper time,  
that can be exhibited, for example, by rewriting  
the following form factor,  
\begin{eqnarray}  
    && \frac1{s (\Box_1- \Box_2) }  
    \Big(f( - s \Box_1)- f( - s\Box_2)\Big)  
\nonumber\\&&\mbox{}  
    = \int_0^1\!\! {\mathrm d}\alpha  
    \alpha (1 - \alpha)  
    \int_0^1\!\! {\mathrm d}\beta    
	\exp \Big( s \alpha(1-\alpha)   
    ((1 - \beta) \Box_1 + \beta \Box_2 ) \Big).   
\end{eqnarray}   
  
Local coefficients of the Schwinger-DeWitt expansion   
\cite{DeWitt-book65,BarVilk-PRep85},  
which is often used in quantum field theory,    
can be easily obtained from  the nonlocal expression  
(\ref{TrK}) by the simple expansion of all form factors  
in powers of the proper time \cite{BGVZ-JMP94-asymp}.

  
\section{One-loop effective action for scalar fields  
coupled to the dilaton in two dimensions}  
\label{result}  
  

The trace of the heat kernel is a classical object, which,   
nevertheless, contains complete information  
about all quantum averages.  
For example, the trace anomaly in two dimensions   
is completely defined by the first Schwinger-DeWitt  
coefficient, $a_1(x)={\mathrm tr}~\hat{P} (x)$  
[the potential term figures here in the form (\ref{potential}),  
rather than in the integrated form (\ref{P})].  
This is a local expression, and   
any derivations of the one-loop effective action  
based just on the coefficient $a_1$    
ignore complex conformally invariant,   
nonlocal structures of the heat kernel.  
Such methods work well in the case of pure $2D$ gravity and give  
the Polyakov effective action,
but they fail in the case of dilaton gravity models.  
However, these procedures still might be valid   
for another field model \cite{KummVass-hepth9811,KummVass-grqc9907}   
discussed  in the closing section of this paper.  
  
In two dimensions, even after subtracting the ultraviolet divergences,
the resulting one-loop renormalized effective action
is not generally defined because of  
bad behavior of the heat kernel trace in  
the large proper time limit (infrared divergence) \cite{CPT2}.   
However, in  our  model (\ref{newmodel})-(\ref{operator})  
we can control the infrared behavior 
using the asymptotic behavior  of the form factors  
\cite{BGVZ-JMP94-asymp},  
\begin{equation}  
	f(-s\Box) = -\frac1s\frac2{\Box}  
	+{\rm O}\left(\frac1{s^2}\right),  
	\hspace{7mm} s\rightarrow\infty   
\end{equation}  
\begin{equation}  
	F(-s\Box_1,-s\Box_2,-s\Box_3) =  
	\frac1{s^2}\Big(\frac1{\Box_1\Box_2}+\frac1{\Box_1\Box_3}+  
	\frac1{\Box_2\Box_3}\Big)  
	+{\rm O}\left(\frac1{s^3}\right),  
\hspace{7mm} s\rightarrow\infty,  
\end{equation}  
and prove the infrared finiteness of $W$.  
Indeed, we find that   
\begin{eqnarray}  
\frac1s{\rm Tr} K(s)&=&{\rm O}\left(\frac1{s^2}\right),  
\hspace{7mm}  s\rightarrow\infty.   
\end{eqnarray}        
and, hence, the proper time integral is convergent at the upper limit.

In order to compute the integral (\ref{efac}) we apply   
the technique of Ref.~\cite{BGVZ-JMP94-asymp}.  
Let us reproduce here some differential equations  
that basic form factors of the nonlocal heat kernel  
satisfy:  
\begin{eqnarray}&&  
	-s\frac{\Box_1\Box_2\Box_3}{D}F(-s\Box_1,-s\Box_2,-s\Box_3)=  
	\frac{\mathrm d}{{\mathrm d} s}
	\Big(sF(-s\Box_1,-s\Box_2,-s\Box_3)\Big)  
\nonumber\\&&\mbox{}  
	+\frac{\Box_1(\Box_3+\Box_2-\Box_1)}{2D}f(-s\Box_1)  
	+\frac{\Box_2(\Box_1+\Box_3-\Box_2)}{2D}f(-s\Box_2)  
\nonumber\\&&\mbox{}  
	+\frac{\Box_3(\Box_1+\Box_2-\Box_3)}{2D}f(-s\Box_3), \label{difeq1}  
\\[3mm]&&  
	\frac{f(-s\Box)-1}{s\Box}=  
	\frac{\mathrm d}{{\mathrm d} s}  
	\left(-\frac{2}{\Box}f(-s\Box)\right)  
	+\frac12f(-s\Box),                  \label{difeq2}  
\\[3mm]&&  
	\frac{f(-s\Box)-1-\frac16s\Box}{(s\Box)^2}=  
	\frac{\mathrm d}{{\mathrm d} s}  
	\left(-\frac{2}{3\Box}  
	\frac{f(-s\Box)-1}{s\Box}  
	-\frac{1}{3\Box}f(-s\Box)\right)  
\nonumber\\&&\mbox{}  
	+\frac{1}{12}f(-s\Box),    \label{difeq3}  
\end{eqnarray}  
where $D$ is the expression  
\begin{equation}  
	D={\Box_1}^2+{\Box_2}^2+{\Box_3}^2-2\Box_1\Box_2  
	-2\Box_1\Box_3-2\Box_2\Box_3. \label{DD}  
\end{equation}  
  
Applying these relations to the heat kernel (\ref{HK})   
we can present some part of it in the form  
of a total derivative over the proper time:  
\begin{eqnarray}  
    \frac1s{\rm Tr} K(s) &=&  
    \frac1{4\pi}\int\! {\mathrm d} x\, g^{1/2}\,    
\left\{  
    \frac{\mathrm d}{{\mathrm d} s}   
    \Big[  
      f_1 (s| \Box_2) R_1 R_2  
    + f_2 (s| \Box_2)   
    \Big( \Box \phi_1 -   
    	(\nabla \phi_1)^2\Big) R_2    
\right.\nonumber\\&&\mbox{}  
    + N_1 (s|\Box_1,\Box_2,\Box_3) R_1 R_2 R_3    
    + N_2 (s|\Box_1,\Box_2,\Box_3) R_1 R_2 (\Box \phi_3 )  
\nonumber\\&&\mbox{}  
    + N_3 (s|\Box_1,\Box_2,\Box_3) 	R_1 (\Box \phi_2)(\Box \phi_3)    
    + N_4 (s|\Box_1,\Box_2,\Box_3) (\Box \phi_1)(\Box \phi_2)(\Box \phi_3)  
\Big]   
\nonumber\\&&\mbox{}   
    + \frac1{s^2}   
       +(\Box \phi) \frac{1}{s\Box}  (\Box \phi)   
    +  h (s| \Box_2)   
    (\Box \phi_1) (\Box \phi_2)  
\nonumber\\&&\mbox{}   
    + H (s| \Box_1,\Box_2,\Box_3)  
    R_1 (\Box \phi_2)(\Box \phi_3)  
    +{\mathrm O}[\Re^4]  
 \Big\},     \label{totalD}  
\end{eqnarray}   
where   
\begin{eqnarray}  
    f_1(s| \Box_2) &=&  
    \frac{1}{\Box_2}\left(\frac18f(-s\Box_2)-\frac14  
    \frac{f(-s\Box_2)-1}{s\Box_2}\right),   
\\[3mm]\mbox{}  
    f_2(s| \Box_2)&=&  
    \frac{1}{\Box_2} f(-s\Box_2),  
\end{eqnarray}   
and  
\begin{eqnarray}                              
    h (s| \Box_2) &=&   
    \frac12  f(-s {\Box_2}),  
\\  
    H (s| \Box_1,\Box_2,\Box_3)&=&  
    -\frac{1}{2\Box_1}\frac{1}{\Box_2 - \Box_3}  
    (\Box_2 f(-s\Box_2) -  \Box_3 f(-s\Box_3)). \label{H}  
\end{eqnarray}   
The third-order form factors $N_i (s|\Box_1,\Box_2,\Box_3)$   
may be found  in Appendix B.  
As can be seen from (\ref{totalD}), not all of the terms in   
${\mathrm Tr} K(s)$ admit the form of a total derivative.  
As a result, in contrast to the Polyakov action   
\cite{Polyakov-PLB81,BGVZ-JMP94-asymp},  
the effective action (\ref{efac})
in two dimensions depends on the ultraviolet   
cutoff parameter $\mu$.

For the sake of convenience we split the total renormalized 
effective action into two parts,
	$    W_{\mathrm ren}= W_{\mathrm fin} + W_{\mu}$.  
	$W_{\mathrm fin}$ 
is a part defined by the total  
derivative terms of Eq.~(\ref{totalD}) while  
	$W_{\mu }$ 
is defined by the rest (including any higher order terms).   
The calculation of the finite  
terms becomes trivial as we perform the proper time integration.  
Using Eqs.~(\ref{difeq1}), (\ref{difeq2}) again we can check that  
the form factors $f_i$ and $N_i$ vanish at  
the upper limit, \hbox{$s\rightarrow\infty$}. Thus,  
\begin{eqnarray}  
W_{\mathrm fin} &=&  
    \frac1{8\pi}\int\! {\mathrm d}^2 x\, g^{1/2}\,  
    \Big\{   
	f_1(s=0) R_1 R_2   
    + f_2(s=0)   
    \Big( \Box \phi_1 -   
    (\nabla \phi_1)^2 \Big) R_2   
\nonumber\\&&\mbox{}  
    + N_1^{\mathrm sym}(s=0) R_1 R_2 R_3   
    + N_2^{\mathrm sym}(s=0) R_1 R_2 (\Box \phi_3)   
\nonumber\\&&\mbox{}   
    + N_3^{\mathrm sym}(s=0) R_1 (\Box \phi_2) (\Box \phi_3)   
\nonumber\\&&\mbox{}   
    + N_4^{\mathrm sym}(s=0) (\Box \phi_1) (\Box \phi_2)(\Box \phi_3)  
\Big\},  
\end{eqnarray}  
where  
$N_i^{\mathrm sym} (i= 1, \ldots  4)$  
are symmetrized in their arguments, $\Box_1,\Box_2,\Box_3$,  
according to the symmetries of the tensor structures  
they are acting on. All of the third-order ($\Re^3$) contributions
to the heat kernel trace vanish, leaving the expression  
\begin{equation}  
	W_{\mathrm fin} =\frac1{96\pi}\int\!   
    {\mathrm d}^2 x\, g^{1/2}\,   
	\left\{  
	R\frac{1}{\Box}R   
	- 12 (\nabla \phi)^2 \frac{1}{\Box} R   
	+ 12  \phi R   
\right\}. \label{Wfin}  
\end{equation}  
This part coincides with the usual form of  
the one-loop effective action derived by  
the integration of the trace anomaly   
\cite{KummLieblVass-1997,MukhWipfZel-PLB94,%
Dowker-CQG98,ChibaSiino-MPLA97}.

To treat the remaining terms of the effective action
(when the zeroth order term is already subtracted \cite{FradVilk-LNC77})   
we apply the proper time cutoff regularization,    
\begin{eqnarray}  
    W_{\mu } &=&  
    -\frac1{8\pi}\int\! {\mathrm d}^2 x\, g^{1/2}\,   
    \int^{L}_{1/{\mu^2}} {\mathrm d} s  
    \Big\{  
    \frac1s \phi \Box \phi   
    +  h ( s | \Box_2)   
    (\Box \phi_1) (\Box \phi_2)  
\nonumber\\&&\mbox{}   
    + H (s | \Box_1,\Box_2,\Box_3)  
    R_1 (\Box \phi_2)(\Box \phi_3)  
     +{\mathrm O}[\Re^4]   \Big\},   \label{Wdiv}                 
\end{eqnarray}     
where infrared and ultraviolet   
cutoff parameters  are  introduced   
correspondingly at the upper and lower limits of the proper time integral   
in order to single out terms of this integral   
that are apparently divergent.
However, the integral as a whole   
appeared to be infrared finite and independent of the parameter  
$L$, as will be demonstrated in a moment.  
As far as ultraviolet divergences are  concerned,  
only the first, local, term of integral (\ref{Wdiv})  
 is divergent when $\mu\rightarrow \infty$.  
Then, the key element of our  computations is the  integral:   
\begin{equation}  
	 \int_0^L\! {\mathrm d} s  
	f (-s \Box) =  
	\int_0^1\! {\mathrm d}\alpha   
	\frac{ {\mathrm e}^{\alpha(1-\alpha)L \Box}   
	- 1}{\alpha(1-\alpha)\Box}  
	= - \frac{2}{\Box}\,   
	\Big(\ln (- L \Box) + {\bf C} \Big),  
\end{equation}    
where ${\mathbf C}$ is the Euler constant.  
The dependence on the infrared cutoff parameter $L$   
in first two terms of integral (\ref{Wdiv})   
cancels, as does the $L$~-dependence in   
the form factor (\ref{H}) of the third term.  
The resulting expression reads  
\begin{equation}  
	W_{\mu } =  
	\frac1{8\pi}\int\! {\mathrm d}^2 x\, g^{1/2}\,  
\left\{   
	(\Box  \phi) \ln (- \frac{\Box}{\mu^2}) \phi    
	 -  \frac{ \ln ( {\Box_2}/{\Box_3}) }  
	{ ({\Box_2}- {\Box_3})} \frac1{\Box_1}  
	R_1 (\Box \phi_2) (\Box \phi_3)   
	+ {\mathrm O}[\Re^4]   
\right\}. 				\label{Wdiv2}  
\end{equation}  
It should be emphasized that the nonlocal form factors of   
Eq.~(\ref{Wdiv2}) can be used for physical applications  
only when expressed in terms of the Green function, e.g.,
after converting them into the mass spectral integrals  
\cite{CPT2,CPT3},  
\begin{equation}  
	- \ln (- \frac{\Box}{\mu^2}) =   
	\int_0^{\infty} {\mathrm d} m^2  
\left(  
	\frac{1}{m^2 - \Box} -  
	\frac{1}{m^2 - \mu^2}  
\right),  			\label{log}  
\end{equation}  
\begin{equation}  
	- \frac{\ln ( {\Box_1}/{\Box_2} )}{\Box_1 - \Box_2} =   
	\int_0^{\infty} {\mathrm d} m^2  
	\frac{1}{m^2 - \Box_1}\,   
	\frac{1}{m^2 - \Box_2}, \label{log12}  
\end{equation}  
where $1/(m^2- \Box)$ is the massive Euclidean Green function  
with zero boundary conditions at spacetime infinity.

Combining the two pieces,  
$W_{\mathrm fin}$ and  $W_{\mu }$,   
we get the final result for the renormalized  
effective action  
\begin{eqnarray}  
	W_{\mathrm ren}&=&\frac1{96\pi}\int\! {\mathrm d}^2 x\, g^{1/2}\,   
	\left\{  
	R\frac{1}{\Box}R   
	- 12 (\nabla \phi)^2 \frac{1}{\Box} R   
	+ 12  \phi R   
	 + 12 (\Box \phi) \ln (- \frac{\Box}{\mu^2}) \phi    
	\right.  
	\nonumber\\&&\mbox{}  
	\left.  
	 - 12 \frac{ \ln ( {\Box_2}/{\Box_3}) }  
	{ ({\Box_2}- {\Box_3})} \frac1{\Box_1}  
	R_1 (\Box \phi_2) (\Box \phi_3)   
	+ {\mathrm O}[\Re^4]   
    \right\}. \label{main}  
\end{eqnarray}  
Equation (\ref{main}) is one of  the main results of this paper.  
This effective action is covariant by construction.  
It evidently reproduces the  conformally anomalous part  
\cite{MukhWipfZel-PLB94}  
and unambiguously fixes new conformally invariant terms that were  
not derived previously in the literature.  
It can be seen explicitly from  
(\ref{main}) that, with an exception for   
the anomalous $R \phi$ term,   
terms of the first and third orders   
in the dilaton field are absent from 
the conformally invariant part of this one-loop effective action.  
This plausibly indicates that all higher-order terms  
are even in powers of the dilaton. This conjecture can be tested  
using the higher-order perturbation expansions for the heat kernel   
in flat space found in Refs.~\cite{WilkFujOsb-PRA81,FliegNSS-Winn94}.    
  
In agreement with Ref.~\cite{LombMazzRusso-PRD99},  
we see that the $\phi^2$ terms of the effective action   
are nonlocal, and do not have the local form derived in   
\cite{KummVass-hepth9811,KummVass-grqc9907}.  
We should stress that these terms are infrared finite. 
Only the ultraviolet regularization parameter $\mu$ 
enters the answer, and  not the infrared cutoff as  
was suggested in \cite{LombMazzRusso-PRD99}.  
  
The form of the last term  in Eq.~(\ref{main}) 
is different from the one in Ref.~\cite{LombMazzRusso-PRD99}  
because the covariantization procedure   
of nonlocal terms used there is incorrect. 
To check our result we can perform the opposite operation,  
namely, to sum the series of the terms quadratic in the dilaton  
into a flat space object.  
To do so we take the expression (\ref{Wdiv2})  
in a flat space assuming that the original metric  
is related to the flat space one $\bar{g}_{\mu\nu}$  
via a conformal factor:  
$g_{\mu\nu}={\mathrm e}^{- 2 \sigma} \bar{g}_{\mu\nu}$.  
To return to the original metric we  
have to use the equation for the variation 
of the effective  action form factor \cite{BGVZ-NPB95},  
\begin{equation}   
    \int \! {\mathrm d}^2 x \, g^{1/2}   
    \delta_{\sigma} \left( \ln \left(- 
\frac{{{\bar{\phantom{'}\Box}\phantom{'}}}_2}{\mu^2}\right) \right) 
	\Re_1  \Re_2  
    = \int \! {\mathrm d}^2 x \, g^{1/2}  
	\frac{\ln ( {\Box_1}/{\Box_2} )}{\Box_1 - \Box_2}  
	\delta_{\sigma} ( \Box_2 )      
	\Re_1 \Re_2  + {\mathrm O}[\Re^3],	\label{ffvar}  
\end{equation}   
where ${{\bar{\phantom{'}\Box}\phantom{'}}}$ is  
defined for the flat-space metric  
$\bar{g}_{\mu\nu}$.  
Eq.~(\ref{ffvar}) follows directly from the rule of variation  
of the Euclidean Green function \cite{DeWitt-book65},  
\begin{equation}   
    \delta_{\sigma} \frac{1}{m^2 - \Box}=  
    \frac{1}{m^2 - \Box}\,   
    \delta_{\sigma} ( \Box )\,  
    \frac{1}{m^2 - \Box},	\label{Green-var}  
\end{equation}   
and Eqs.~(\ref{log}), (\ref{log12}).  
We know that  $\delta_{\sigma} ( \Box ) = - 2 (\delta \sigma) \Box$,
where $\delta \sigma = \sigma$,  
and upon substituting the nonlocal curvature expression  
for the conformal factor,  
$\sigma (g) = - \frac12 \frac{1}{\Box} R$,  
into (\ref{ffvar}) one can see that it becomes nothing but   
the second term of Eq.~(\ref{Wdiv2}).  
As a result, it is possible   
to rewrite the terms quadratic in the dilaton field   
in a flat spacetime form,  
\begin{equation}  
	W_{\mu } =   
	\frac1{8\pi} \int\!{\mathrm d}^2 x\,     
\left\{  
	({{\bar{\phantom{'}\Box}\phantom{'}}} \phi )   
	\ln(- \frac{{{\bar{\phantom{'}\Box}\phantom{'}}}}{\mu^2}) \phi   
	+ {\mathrm O} [\phi^4]   
\right\}.  \label{answer}  
\end{equation}  
When expanded in powers of the curvatures,  
the effective action (\ref{answer})  again
becomes an infinite  series.

  
\section{Partial summation of the dilaton effective action}  
  
  
In the previous sections  we have derived the one-loop effective action  
as a perturbation series in powers of the dilaton  
field $\phi$. So far no explicit form  
 of the dilaton field was assumed,  
and we could rewrite this expansion in terms of the potential $P(\phi)$.  
But instead of doing a perturbative expansion  
using the potential as a small parameter, we perform a partial  
summation of the effective action and obtain  
a result, which is nonperturbative in terms of $P$.  
To simplify calculations we work in a flat spacetime  
in the present and following sections.  
In other words, we perform all operations only with  
the conformal part of $W_{\mathrm ren}$, Eq.~(\ref{answer}),   
and we can restore the whole covariant result,  
expressed in terms of $R$ and $P$,  
at the end of the derivations. For the sake of convenience,  
we keep here the covariant notation $\Box$ instead of  
the flat-space one ${{\bar{\phantom{'}\Box}\phantom{'}}}$.  
  
Let us  begin with the equation for the potential $P$,  
\begin{equation}  
    {P}= \Box \phi -   
    (\nabla_{\mu} \phi) (\nabla^{\mu} \phi).   
    \label{potentialP}  
\end{equation}  
We rewrite this equation as a linear differential equation  
by substituting the following ansatz for the dilaton field  
\cite{MukhWipfZel-PLB94},  
\begin{equation}  
	\phi = - \ln \Omega. \label{ln_omega}  
\end{equation}  
The solution of the resulting equation on $\Omega$,   
\begin{equation}  
    (\Box  + P ) \Omega =0,   
    \label{potentialP2}  
\end{equation}  
reads  
\begin{equation}  
     \Omega = 1 - \frac{1}{\Box + P} P,    
    \label{omega}  
\end{equation}  
where the boundary condition $\Omega=1$   
at $|x| \rightarrow \infty$ is assumed.  
In principle, more general solutions containing  
zero modes,  $ \Box \Omega_0 =0$,  
are allowed, but   
the requirement of covariant perturbation theory \cite{CPT2}  
that all  background fields  
including the dilaton field (\ref{ln_omega})  
vanish at spacetime infinity puts $\Omega_0 = 1$.  
  
The effective action, which is known up to the third order  
in the dilaton field, now can be written in  
a nonperturbative form by inserting (\ref{ln_omega}) and (\ref{omega})   
into Eq.~(\ref{answer}). The result reads  
\begin{equation}  
	W_{\mathrm \mu} =  
	\frac1{8\pi}\int\!{\mathrm d}^2 x\,   
	(\Box \phi )\, \ln(-\frac{\Box}{\mu^2})  
	\, \phi,  \label{resummedW}   
\end{equation}  
where  
\begin{equation}  
     \phi (x)  = - \ln \left( 1 - \frac{1}{\Box + P} P \right).   
    \label{phi-omega}  
\end{equation}  
Thus, we obtain a partially summed form of  
the one-loop effective action.   
This summation is partial, because  
not all higher-order terms are included   
in (\ref{resummedW}), (\ref{phi-omega})   
but only those containing the form factor  
$\ln(- \Box/{\mu^2})$.   
Such a  summation with help of a new   
auxiliary scalar field, which is expressed in a    
nonlocal way through the perturbation (curvature)  
\cite{FradkVilk-PLB78},  
is very similar to the summation  
of the Ricci scalar terms in the  4D covariant  
effective action performed in   
Ref.~\cite{MirzVilkZhyt-PLB95}.  
  
To reproduce the perturbation series we first obtain  
the expansion of  $\Omega$,  
\begin{equation}  
   	\Omega (x) =   
		1-  \frac{1}{\Box}P   
	 + \frac{1}{\Box} \Big( P \frac{1}{\Box} P \Big)      
	+ {\mathrm O}[P^3],    
\end{equation}  
which gives us an approximation for the dilaton   
\begin{equation}
   \phi (x) = \frac{1}{\Box} P 
	+ \frac12 \Big( \frac{1}{\Box} P \Big) \Big( \frac{1}{\Box} P \Big) 
	- \frac{1}{\Box} \Big( P \frac{1}{\Box} P \Big) 
	+ {\mathrm O}[P^3].
\end{equation}
This series obviously coincides with 
an iterative solution of Eq.~(\ref{potentialP}).
The perturbative expansion of (\ref{resummedW}) reads
\begin{eqnarray}
	W_{\mathrm \mu} & = &
	\frac1{8\pi}\int\!{\mathrm d}^2 x\, 
\left\{ 
	P \, \ln(-\frac{\Box}{\mu^2})
	\, \Big( \frac{1}{\Box} P \Big) \right.
	+ \frac12 P \, \ln(-\frac{\Box}{\mu^2})
	\, \left( \Big( \frac{1}{\Box} P \Big) 
		\Big( \frac{1}{\Box} P \Big) \right) 
\nonumber\\ &&\mbox{}
	+ \frac12 \Box \left( \Big( \frac{1}{\Box} P \Big) 
		\Big( \frac{1}{\Box} P \Big) \right) 
	\, \ln(-\frac{\Box}{\mu^2})
 	\, \Big( \frac{1}{\Box} P \Big) 
	- \Big( P \frac{1}{\Box} P \Big) 
	\, \ln(-\frac{\Box}{\mu^2})
 	\, \Big( \frac{1}{\Box} P \Big) 
\nonumber\\ &&\mbox{}
 	- P \, \ln(-\frac{\Box}{\mu^2}) \left.
	\frac{1}{\Box} \Big( P \frac{1}{\Box} P \Big) 	
	+ {\mathrm O}[P^4]
\right\}.  \label{W_P} 
\end{eqnarray}
The leading term of this expansion is apparently 
similar to the expression obtained in   
Ref.~\cite{LombMazzRusso-PRD99}  
with the reservation of a different meaning for   
the regularization parameter $\mu$.

  
\section{Dilaton effective action at finite temperature}  
  

An obvious use for the obtained one-loop effective action  
is its application to the calculation of the stress tensor.   
So far we have worked in a Euclidean spacetime.  
According to rules of covariant perturbation theory,  
one makes the transition to Minkowski spacetime  
only after deriving quantum averages and currents from 
the Euclidean effective action \cite{CPT1}.   
Similarly, in the calculation of the energy-momentum tensor   
for quantum fields in a black hole background  
boundary conditions corresponding to  
the Unruh, Boulware, or Hartle-Hawking vacua  
are to be specified after the variation over the metric.   
Authors of \cite{BalbFabbri-PRD99,BalbFabbri-grqc99}   
completed this procedure by making  
the effective action local by introducing  auxiliary fields  
and imposing the proper boundary conditions on them.  
  
The other way to introduce boundary conditions  
corresponding to the Hartle-Hawking vacuum  
is to consider the field system at some fixed temperature  
$T=1/\beta$. This is relatively easy to do,  
because the Killing vector always exists in two dimensions,
in contrast to higher dimensions   
\cite{DowkKenn-JPA78,GusZel-CQG98,GusZel-PRD99}.  
Therefore, without losing generality,  
we can make a conformal transformation  
to a flat space where the Euclidean time is periodic,  
i.e., the flat spacetime has the topology of a cylinder.  
In our new flat space the anomaly-generating part   
of the effective action (\ref{Wfin}) vanishes,  
so we deal only with the conformally invariant part  
(\ref{Wdiv2}). This is what one would expect,  
because the anomalous part does not depend on temperature.  
In our treatment of the finite-temperature  
effective action for scalar fields coupled to the dilation  
we will follow the computational scheme  
of Refs.~\cite{GusZel-CQG98,GusZel-PRD99}.  
For general notions of finite temperature field theory  
we refer to \cite{DowkKenn-JPA78},  
and references on some earlier works   
can be found in \cite{GusZel-PRD99}.  
  
Let us start with the flat-space limit of the trace of the heat kernel  
(\ref{TrK}),  
\begin{eqnarray}  
    {\mathrm Tr} K(s) &=&  
    \frac1{(4\pi s)^{D/2}}\int\! {\mathrm d}^D x\,   
    \left\{  
  	1  
    +  s P  
    + s^2 \frac12  P f(-s {\Box}) P   
    + {\mathrm O} [P^3]  
\right\}.  
\end{eqnarray}  
Here we restrict our consideration to terms of the second order,  
because it gives  the first nonlocal contribution to 
the finite-temperature effective action $W^{\beta}$.  
Using the form (\ref{potential}) for the potential term 
we rewrite this heat kernel in terms of the dilaton field,  
\begin{equation}  
   {\mathrm Tr} K(s) =
    \frac1{(4\pi s )^{D/2}}\int\! {\mathrm d}^D x\,   
    \left\{  
  	1  
    + s \phi \Box \phi   
    + s^2 \frac12  (\Box {\phi}) f(-s {\Box}) (\Box {\phi})   
    + {\mathrm O} [ \phi^3 ]  
\right\}.  
\end{equation}

We are calculating $W^{\beta}$ in a way similar to   
the zero temperature case (\ref{efac}),  
\begin{equation}   
	 W^{\beta}=   
	\mbox{}-\frac1{2}    
	\int_0^{\infty} \frac{\mathrm{d} s}{s}   
\left(  
	{\mathrm{Tr}}  K^{\beta} (s)  
	- {\mathrm Tr} K (s)|_{\phi=0}  
\right), \label{efacT}  
\end{equation}   
where we subtract the zeroth-order term of the zero-temperature  
${\mathrm Tr} K (s)$ from the heat kernel at 
some  finite temperature $1/\beta$.  
It is well known \cite{DowkKenn-JPA78} that one can express   
the heat kernel at finite temperature  
as an infinite sum of the zero temperature heat kernels  
at separated points $x$ and $x'$,  
\begin{equation}  
	K^{\beta}(s|\tau, \mbox{\boldmath $x$}; \tau',   
	\mbox{\boldmath $x$}')  
	=\sum_{n=-\infty}^{\infty}   
	K(s|\tau, \mbox{\boldmath $x$}; {\tau}'+\beta n,   
	\mbox{\boldmath $x$}'),  \label{sumHK}        
\end{equation}  
where $\tau$ is the Euclidean time and $\mbox{\boldmath $x$}$
are the spatial coordinates.
Then, the two-dimensional ${\mathrm Tr} K^{\beta}$ can be found via  
the heat kernel in one dimension,  
\begin{eqnarray}   
	{\mathrm  Tr} {K}^{\beta}(s)&=&    
	\theta_3    
	\Big(0,   
 	 {\mathrm e}^{-\frac{\beta^2}{4s}}    
	\Big)    
	\frac{\beta}{(4\pi s)^{1/2}}    
	\int {\mathrm  d}^1 x \,     
	{\mathrm{tr}}\,  K^{(1)}   
	(s| \mbox{\boldmath $x$}, \mbox{\boldmath $x$}),  
\end{eqnarray}   
where we have introduced the Jacobi  theta function ,   
\begin{equation}  
	\theta_3 (a,b)  \equiv   \sum_{n=-\infty}^{n=\infty}   
	{{\mathrm{e}}  }^{2 n a{\mathrm{i}}} b^{n^2}. \label{theta3}  
\end{equation}

The zeroth-order term of the effective action requires  
special treatment. First of all, it is removed by   
the renormalization procedure in the zero temperature QFT  
[see Eq.~(\ref{efacT})],  
which amounts to subtracting the $n=0$ term from 
the sum (\ref{sumHK}).  
Straightforward computation of the proper time integral  
(\ref{efacT}) with the subsequent summation  
over $n$ gives us a numerical coefficient in front of  
the $1/\beta$ term, which figures in the final  
expression (\ref{Wbeta}) below.  
  
The dilaton-dependent part of ${W}_{\beta}$  
can be computed after the Poisson re-summation,  
\begin{equation}  
    \sum_{n=-\infty}^{\infty}   
    {\mathrm{e}}^{- \frac{\beta^2}{4 s}n^2}  =   
    \frac{\sqrt{4\pi s}}{\beta}  
    \sum_{ k = -\infty}^{\infty}       \label{Poisson}  
    {\mathrm{e}}^{- \frac{4 \pi^2 s }{ \beta^2 } k^2 }.    
\end{equation}    
The term $k=0$ of a new sum over $k$    
corresponds to the high temperature limit 
$T \rightarrow \infty$ ($\beta = 0 $),  
and it is ultraviolet finite.   
The rest of the $k$-sum corresponds to a single term  
$n=0$ ($T=0$) of the original $n$-sum, thus, it   
needs to be regularized. All of this follows from   
the fact that ultraviolet counterterms introduced  
in a field theory at zero temperature are sufficient  
for renormalization of the finite temperature field theory.  
Here we use the zeta function regularization,  
\begin{equation}   
	 W^{\beta}_{k \neq 0}=   
	\mbox{}-\frac12 \frac{\partial}{\partial\epsilon}   
	\left[   
	\frac{\mu^{2\epsilon}}{\Gamma(\epsilon)}   
	\int_0^{\infty}\! \frac{{\mathrm{d} }s}{s^{1-\epsilon}}\,    
	{\mathrm Tr}  {K}^{\beta}|_{k\neq 0}(s)  
	\right]_{\epsilon=0},  
\end{equation}             
where $\epsilon$ is a small positive parameter and  
$\Gamma$ is the gamma function.  
  
On the other hand, the sum over $k$ is infrared finite,  
with an exception for the $k=0$ term.  
Infrared divergences appearing in different  
orders of the perturbation theory in $P$ are,  
in fact, artificial and disappear in the final result  
expressed in terms of $\phi$,   
but we need to introduce an auxiliary mass  
to treat them at intermediate stages,  
\begin{equation}   
	 W^{\beta}_{k=0}=   
	\mbox{}-\frac12   
	\int_0^{\infty}\! \frac{{\mathrm{d} }s}{s}\,    
	 {\mathrm e}^{-s m^2}\Big(   
	{\mathrm Tr}  {K}^{\beta} (s)|_{k=0}  
	- {\mathrm Tr} K (s)|_{\phi=0} \Big)  
	\Big|_{m^2=0}.  
\end{equation}   
One can observe that the infrared poles in $ W^{\beta}_{k=0}$,   
which come from the local $P$ and nonlocal  $P^2$   
contributions, mutually cancel.  
  
The final result reads  
\begin{eqnarray}  
	{W^{\beta}_{\mathrm ren}} &=&  
        - \int\! {\mathrm d} x\,   
        \left\{  
        \frac{\pi}{12 \beta} +   
        \frac{\beta}{8 \pi}  
    (\Box \phi)   
\left[   
    2 \ln \Big(  \frac{\beta \mu}{4 \pi} \Big)      
\right. \right.  
\nonumber\\ && \mbox{}  
    - \Psi \Big( \frac{i \beta \sqrt{- \Box} }{4 \pi} \Big)   
\left. \left.    
    - \Psi \Big( - \frac{i \beta \sqrt{- \Box} }{4 \pi} \Big)  
    \right]  \phi  
	+ {\mathrm O} [ \phi^3 ]    
\right\},     			\label{Wbeta}  
\end{eqnarray}  
where $\Psi$ is the psi function.  
The obtained finite temperature effective action   
is infrared finite but depends on the ultraviolet   
regularization parameter $\mu$.  
Even though arguments of the psi functions are imaginary,  
the combination of $\Psi$'s in Eq.~(\ref{Wbeta})  
is equivalent  to the real part of $\Psi$. 
The thermal form factor found is a smooth function of the inverse
temperature $\beta$, and its different asymptotics can be easily analyzed.

We emphasize that expression (\ref{Wbeta}) is valid  
for arbitrary temperature, and high and low temperature   
asymptotics can be derived from it.  
The most popular expansion in thermal field theory  
is the high temperature one, $\beta \rightarrow 0$.  
We found that this  expansion admits a local form,  
\begin{eqnarray}  
	\frac{ W^{\beta}_{\mathrm ren} }{\beta} &=&  
        - \int\! {\mathrm d} x\,   
\left\{  
        \frac{\pi}{12 \beta^2} +   
        \frac{1}{4 \pi}  
    \Big[   
    \ln \Big(  \frac{\beta \mu}{4 \pi} \Big)  
    + {\mathbf C}   
    \Big] \phi \Box \phi   
\right.  
\nonumber\\ && \mbox{}   
    + \frac{\zeta (3)}{64 \pi^3} \beta   
    (\Box \phi) (\Box \phi)       
    + \frac{\zeta (5)}{1024 \pi^5} \beta^3   
    (\Box \phi) (\Box^2 \phi)   
\nonumber\\ && \mbox{}   
    + \left. \frac{\zeta (7)}{16384 \pi^5} \beta^5   
    (\Box \phi) (\Box^3 \phi)   
	+ {\mathrm O} [ \phi^3 ] + {\mathrm O} [\beta^8]  
\right\},  
	\ \ \ \beta \rightarrow 0.  
\end{eqnarray}  
This series has a form similar to high  
temperature expansions in four dimensions  
\cite{DowkKenn-JPA78,GusZel-PRD99}.  
     
The other important limit is, of course, the low temperature  
asymptotic,   
\begin{eqnarray}  
	\frac{ W^{\beta}_{\mathrm ren} }{\beta} &=&  
    	\int\! {\mathrm d} x\,     
\left\{   
	\frac{1}{8 \pi}  
    	(\Box \phi )  \ln \Big(-\frac{\Box}{\mu^2}\Big) \phi  
	- \frac{1}{\beta^2} \frac{\pi}{12}  
	(1 + 4 \phi^2 )   
\right.  
\nonumber\\ && \mbox{}  
	- \frac{1}{\beta^4} \frac{8 \pi^3}{15}   
\left.	\phi \Big(\frac{1}{\Box} \phi \Big)  
	+ {\mathrm O} [ \phi^3 ]  
	+ {\mathrm O} [1/\beta^6]  
\right\},  
\ \ \ \beta \rightarrow \infty.  
\end{eqnarray}  
Restoring the effective action to the  
original two-dimensional spacetime form  
one can see that the leading, temperature independent,   
term of this expansion is just $W_{\mathrm ren}$ at zero  
temperature (\ref{Wdiv2}).   

In curved spacetime the total result will be the sum of   
the anomalous part of the effective action (\ref{Wfin}),
and Eq.~(\ref{Wbeta}) with all flat quantities,  
metric and derivatives, being  
expressed in terms of physical (curved spacetime) ones.  
We are not aware of similar results in the literature 
to which one could make any comparisons.

  
\section{Discussion}  
  

We have calculated the one-loop effective action for scalar fields
interacting with a background dilaton field in curved spacetime.
The main results are Eqs.~(\ref{main}) and (\ref{resummedW}) for 
the zero-temperature case and Eq.~(\ref{Wbeta}) 
for the finite-temperature one. 
Strictly speaking, these results are obtained for a generic
dilaton field as long  as   
the corresponding potential $P$ has the form (\ref{potential}), 
and decreases sufficiently rapidly at infinity.
These results are applicable to arbitrary two-dimensional spacetimes 
with the topology of either a disk or a cylinder,
because in two dimensions there is always a Killing vector,
and using nonsingular conformal transformations 
the  problem can be reduced to one for a flat spacetime 
with the corresponding topology.

The difference between our approach and a similar attempt to 
use perturbation theory made in the interesting paper 
\cite{LombMazzRusso-PRD99} is that 
we are able to control the infrared divergences that
appear in two-dimensional calculations. 
We computed  the one-loop effective action as an expansion in powers  
of the dilaton field rather than in the potential term  of 
a differential operator, and proved the infrared finiteness 
of the effective action order by order. 
The resulting covariant nonlocal effective action 
does not depend on the infrared cutoff, and is proved
to be valid up to the third order 
in the spacetime curvature and the dilaton.  
For one particular field model in a two-dimensional flat spacetime
infrared finiteness of the one-loop two-point functions 
was demonstrated in Ref.~\cite{BalKummPigSchwed-PLB92}.   
Lombardo, Mazzitelli and Russo \cite{LombMazzRusso-PRD99} found,
using results of Ref.~\cite{BalKummPigSchwed-PLB92},   
the nonlocal form factor   
entering the conformally invariant part of the 2D effective action. 
Unfortunately, their generalization 
of this result to curved spacetime (covariantization) was not
correct. The correct form of the covariantization procedure  
makes use of Eq.~(\ref{ffvar}).  
By applying it to the terms quadratic in the dilaton  
we summed them up into a single flat-space term (\ref{answer}).  
Then, the dilaton field could be expressed in
terms of the potential term, Eq.~(\ref{resummedW}), and, 
if necessary, be expanded in powers of the potential
and/or curvatures.

Eq.~(\ref{main})  demonstrates that the 
effective action for scalars interacting with the dilaton does not admit  
the exact form proposed by Kummer and Vassilevich 
in  Refs.~\cite{KummVass-hepth9811,KummVass-grqc9907}, but contains
additional conformally invariant terms that are nonlocal.
Here we would like to explain the origin  of this discrepancy.
In order to calculate the one-loop effective action  
corresponding to the bosonic operator (\ref{operator}) 
they substituted the fermionic representation of the determinant of
the operator 
\begin{equation}  
	\hat{F} (\nabla) =  
	\bar{g}^{\mu\nu} ( {\mathcal D}_{\mu} {\mathcal D}_{\nu}  
	+ \hat{1} \partial_{\mu} \partial_{\nu} \phi ), \label{spinor}  
\end{equation}
in a flat spacetime for an original bosonic representation.
Note that we work in the Euclidean signature and 
the derivative ${\mathcal D}_{\mu}$ is defined   
(for the case $\varphi=\psi=\phi$ in the notation   
of \cite{KummVass-hepth9811}) as  
\begin{equation}  
	{\mathcal D}_{\mu} = \partial_{\mu}  
	- \gamma^5 \epsilon^{\nu}_{\ \mu}   
	 \partial_{\nu} \phi,  
\end{equation}  
Here $\gamma^5$ belongs to  the algebra of gamma matrices  and   
has the property $(\gamma^5)^2= \hat{1}$. 
This is an absolutely legitimate procedure as long as one 
deals with formal definitions of 
the operator  determinants in flat spacetime (\ref{trln}).
But the renormalization procedure generically breaks 
the identity of these two representations,
hence, the corresponding renormalized 
effective actions do not coincide. 
To check this fact by an explicit calculation,
one can  note that the operator (\ref{spinor}) 
also belongs to the class of operators of the type (\ref{op}).
Therefore, one can directly apply the covariant perturbation theory  
to this model. Let us write down the corresponding potential term,
\begin{equation}  
	\hat{P} = \hat{1} {{\bar{\phantom{'}\Box}\phantom{'}}} \phi,  
\end{equation}  
and the commutator curvature,  
\begin{equation}  
	\hat{\mathcal R}_{\mu\nu}  
	= \gamma^5 ( \epsilon^{\alpha}_{\ \mu}   
	 \partial_{\nu} \partial_{\alpha} \phi  
	- \epsilon^{\alpha}_{\ \nu}   
	 \partial_{\mu} \partial_{\alpha} \phi ).  
\end{equation}  
We performed the computation of the one-loop effective action  
along the lines of the calculational scheme   
of sections~\ref{computation} and \ref{result} and using the tables  
of form factors of Ref~\cite{CPT4}.  
We found that the terms of third order in $\phi$,  
which in this case is equivalent to order $\Re^3$,   
vanishes, and the  $\phi^2$ terms have no infrared  singularity. 
The final answer for the effective action corresponding to
the operator (\ref{spinor}) is local and finite:  
\begin{equation}  
	W_{\mathrm ren} =   
	- \frac1{8\pi} \int\!{\mathrm d}^2 x\,     
	\phi  {{\bar{\phantom{'}\Box}\phantom{'}}} \phi   
	+ {\mathrm O} [\phi^4]. \label{vassil}  
\end{equation}  
Indeed, this expression is in good agreement with 
Eq.~42 of Ref.~\cite{KummVass-hepth9811} 
(the opposite sign is attributed to the Euclidean spacetime signature). 
However, as one can see, it is completely  
different from the effective action for the original bosonic model found
in section~\ref{result} above. Technically, this stems from the fact that
in order to remove the ultraviolet divergences one should subtract 
the first two terms of the Schwinger-DeWitt expansion of 
the  heat kernels.  But the Schwinger-DeWitt 
coefficients for the operators in question are absolutely different,
because their heat kernels are different.
Nevertheless, the remarkable result of Ref.~\cite{KummVass-hepth9811} 
is the exact expression for the effective action 
for the spinor model (\ref{spinor}).
This two-dimensional one-loop effective action is local,   
infrared finite, and does not depend on a regularization parameter. 
In principle, we could start with a covariant version
of the differential operator (\ref{spinor}) and derive
the covariant effective action, but it not necessary
because all conformally invariant terms in that effective action
can be obtained from Eq.~(\ref{vassil}) by covariantization,
while the anomalous part is  proved  to be exact 
\cite{KummVass-hepth9811}.

A few remarks are in order about the  applicability of the
perturbative expansion to physically interesting cases. 
The covariant perturbation theory works well when
derivatives of the background fields are much  bigger
than powers of these fields \cite{CPT2}, 
\begin{equation} 
	\nabla \nabla \Re >> \Re^2. \label{smallparameter}
\end{equation} 
Since the anomalous part (\ref{Wfin}) is exact, we are concerned only
about the conformal terms, (\ref{Wdiv2}) or (\ref{answer}).
For instance, if one wants to use the results above 
to study black hole physics and consider 
the 2D dilaton gravity inspired by the spherical reduction from 
four dimensions, then the dilaton field  is defined as  
$\phi= - \ln r$ (where $4\pi r^2$ is the area of a surface of
the constant radius $r$). 
At infinity the 2D Schwarzschild metric
is asymptotically flat, and the 2D curvature and potential obviously 
satisfy the condition  (\ref{smallparameter}). As for the vicinity
of the horizon, note that $W_\mu$ in Eq.~(\ref{answer})
is expressed in terms of the flat spacetime operator 
${{\bar{\phantom{'}\Box}\phantom{'}}}$. 
This means that all quantities in Eq.~(\ref{smallparameter}) 
should also be defined in the flat metric $\bar{g}_{\mu\nu}$. 
One can check that the condition $\bar{\nabla} \bar{\nabla} P > > P^2$ 
(in the 'tortoise' coordinates) is also satisfied near the horizon.
This means that our results should be valid  both at infinity and
at the horizon of the 2D dilaton black hole.

Kummer and Vassilevich 
\cite{KummLieblVass-1997,KummVass-hepth9811,KummVass-grqc9907}
considered the case of a more general differential operator,
with the dilaton coupling being defined
in terms of two  arbitrary functions of the dilaton,  
$\varphi (\phi)$ and $\psi (\phi)$. 
The results of this paper are obtained for the simplest
case $\varphi (\phi) = \psi (\phi) = \phi$. 
Strong reasons to study more  general models 
are discussed in Ref.~\cite{KummVass-grqc9907};
here, we just would like to note that such nontrivial dilaton couplings
can be incorporated into our computational method.
This is possible to do, because such models correspond to
the minimal second-order operator
when expressed in terms of new redefined metric and covariant
derivative \cite{KummLieblVass-1997},
but we leave this generalization for some other publication.

In conclusion, we would like to make some general comments 
on the validity of  the anomaly-induced effective actions.
Such effective actions are very attractive,
because they are relatively easy to derive, 
and they have rather simple structures.
However, as was shown above, the anomaly-induced effective actions
are incomplete even for simple two-dimensional field models.
This is also true for four dimensions,
and an interesting  recent work \cite{BalbFabbShapiro-PRL99} would be 
a good illustration. 
The authors applied the nonlocal effective action obtained 
by integration of the four-dimensional 
trace anomaly (the generalized Riegert action) to the study of 
the Hawking radiation. Indeed, this action is nonlocal, 
therefore, it captures some essential features of 
the energy-momentum tensor in curved spacetime.
Nevertheless, as correctly noted in \cite{BalbFabbShapiro-PRL99},
further applications of the Riegert action encounter serious obstacles, 
because it is ill-defined at infinity \cite{ErdmenOsborn-NPB97}.
The four-dimensional nonlocal effective action 
of Barvinsky and Vilkovisky \cite{CPT2,CPT4}
is the only known action with the  right 
properties \cite{ErdmenOsborn-NPB97}.
There are many ways to split  this effective action 
into anomalous and  conformal parts \cite{BarvMirzZhyt-QG95},
and only one of them gives the Riegert action.
In order to obtain physically consistent results
we should always use the entire effective action,
not just its  anomaly-generating part.


\section*{Acknowledgments}

We thank V. Frolov and S. Solodukhin for stimulating  discussions,
and P. Sutton for reading the manuscript.
This work was supported 
by Natural Sciences and Engineering Research Council of Canada.
A. Z. is grateful to the Killam Trust for its financial support.

  
\section*{Appendix A: Third-order form factors for the trace  
of the heat kernel}  
  
  
Below we display form factors of the third order  
in Eq.~(\ref{HK}) of section~\ref{computation}.  
They are expressed in terms of dimensionless  
arguments $\xi_i= -s \Box_i$, $i=1, \ldots, 3$,  
and the denominator,  
\begin{equation}  
	{\Delta} =  
	{\xi_1}^2+{\xi_2}^2+{\xi_3}^2-2\xi_1\xi_2  
	-2\xi_1\xi_3-2\xi_2\xi_3.  
\end{equation}

\begin{eqnarray}  
M_1 (\xi_1, \xi_2, \xi_3) &= & \mbox{}  
    - F(\xi_1, \xi_2, \xi_3)  
    {{ {{\xi_1}^2} {{\xi_2}^2} {{\xi_3}^2}}\over{3 {{\Delta}^3}}}  
\nonumber\\&&\mbox{}  
    - f( \xi_1)\frac{1}{32 {{\Delta}^3} \xi_2}  
    ( {{\xi_1}^6} - 4 {{\xi_1}^5} \xi_2 - 4 {{\xi_1}^5} \xi_3  
     + 3 {{\xi_1}^4} \xi_2  
     \xi_3 \nonumber\\&&\mbox{}+ 24 {{\xi_1}^3} {{\xi_2}^2} \xi_3  
     + 5 {{\xi_1}^4} {{\xi_3}^2} + 24  
     {{\xi_1}^3} \xi_2 {{\xi_3}^2}  
      - 2 {{\xi_1}^2} {{\xi_2}^2} {{\xi_3}^2}   
\nonumber\\&&\mbox{}  
    + 32 \xi_1 {{\xi_2}^3} {{\xi_3}^2}  
     - 25 {{\xi_1}^2} \xi_2 {{\xi_3}^3} - 36 \xi_1  
     {{\xi_2}^2} {{\xi_3}^3} + 5 {{\xi_2}^3} {{\xi_3}^3}   
\nonumber\\&&\mbox{}  
     - 5 {{\xi_1}^2}  
     {{\xi_3}^4} - 9 {{\xi_2}^2} {{\xi_3}^4}  
     + 4 \xi_1 {{\xi_3}^5} + 5 \xi_2  
     {{\xi_3}^5} - {{\xi_3}^6} )\nonumber\\&&\mbox{}  
-\left(\frac{f( \xi_1)-1}{ \xi_1}\right)  
\frac{1}{8 {{\Delta}^2}\xi_2}  
  ({{\xi_1}^4} - 2 {{\xi_1}^3} \xi_3  
 - 12 {{\xi_1}^2}  \xi_2 \xi_3   
\nonumber\\&&\mbox{}- 10 \xi_1 {{\xi_2}^2} \xi_3 +  
   8 \xi_1 \xi_2 {{\xi_3}^2} - 2 {{\xi_2}^2} {{\xi_3}^2}  
   + 2 \xi_1 {{\xi_3}^3}  
\nonumber\\&&\mbox{}  
 +   3 \xi_2 {{\xi_3}^3} - {{\xi_3}^4})  
\nonumber\\&&\mbox{}  
    - \!\left(\frac{f( \xi_1)-1-\frac16 \xi_1}{( \xi_1)^2}\right)\!  
    \frac{3}{8 D \xi_2}  
    ( {{\xi_1}^2} + 4 \xi_1 \xi_2 + \xi_2 \xi_3 - {{\xi_3}^2} )  
\nonumber\\&&\mbox{}  
    + \frac1{\xi_2-\xi_3}\frac{\xi_2}{32\xi_1}  
    \Big(f( \xi_2)-f( \xi_3)\Big)  
\nonumber\\&&\mbox{}  
    +\frac{1}{\xi_2-\xi_3}\frac{\xi_2}{8\xi_1}  
\left(  
    \frac{f( \xi_2)-1}{ \xi_2}  
    -\frac{f( \xi_3)-1}{ \xi_3}  
\right)  
\nonumber\\&&\mbox{}  
    + \frac{1}{\xi_2-\xi_3}\frac{3\xi_2}{8\xi_1}  
\left(  
    \frac{f( \xi_2)-1-\frac16 \xi_2}{( \xi_2)^2}\right.  
\nonumber\\&&\mbox{}  
\left.  
    -\frac{f( \xi_3)-1-\frac16 \xi_3}  
    {( \xi_3)^2}  
\right),  
\\[3mm]  
M_2(\xi_1, \xi_2, \xi_3) &= &   
    F( \xi_1,  \xi_2,  \xi_3)  
    \frac{{\xi_1} {\xi_2} {\xi_3}^2}{D^2}  
\nonumber\\&&\mbox{}  
    -  f(  {\xi_1}) \frac{1}{12 D^2 {\xi_2}}   
  (2 {\xi_1}^4-5 {\xi_1}^3 {\xi_3}  
\nonumber\\&&\mbox{}  
    + 3 {\xi_1}^2 {\xi_2}^2+3 {\xi_1}^2 {\xi_3}^2   
    + {\xi_1} {\xi_3}^3+ 11 {\xi_1} {\xi_2} {\xi_3}^2   
    + 11 {\xi_1} {\xi_2}^2 {\xi_3}  
\nonumber\\&&\mbox{}  
    + {\xi_1} {\xi_2}^3 -  6 {\xi_2}^2 {\xi_3}^2  
    + 4 {\xi_2}^3 {\xi_3} + 4 {\xi_2} {\xi_3}^3  
    - {\xi_2}^4-{\xi_3}^4)  
\nonumber\\&&\mbox{}  
    + f(  {\xi_3}) \frac{1}{12 D^2 {\xi_2}}  
  ( - {\xi_3}^4 + {\xi_1} {\xi_3}^3   
    +  3 {\xi_1}^2 {\xi_3}^2 -  5 {\xi_1}^3 {\xi_3}  
\nonumber\\&&\mbox{}  
    + 2 {\xi_1}^4 + {\xi_2} {\xi_3}^3   
    +  17 {\xi_1} {\xi_2} {\xi_3}^2  
    - 12 {\xi_1}^2 {\xi_2} {\xi_3}   
\nonumber\\&&\mbox{}  
    - 6 {\xi_1}^3 {\xi_2}   
    + 17 {\xi_1} {\xi_2}^2 {\xi_3}   
    + 4 {\xi_1}^2 {\xi_2}^2)  
\nonumber\\&&\mbox{}  
    -   \frac{ f(  {\xi_1}) -1 }{  {\xi_1}}  
    \frac{{\xi_1}}{2 D {\xi_2}}({\xi_1} - {\xi_2} - {\xi_3})  
\nonumber\\&&\mbox{}  
    + \frac{ f(  {\xi_3}) -1 }{  {\xi_3}}  
\frac{1}{ 2 D  {\xi_2}}( {\xi_1}^2 - {\xi_1} {\xi_3} -   
    3 {\xi_2} {\xi_3} - {\xi_1} {\xi_2})  
\nonumber\\&&\mbox{}  
    + \frac{1}{\xi_2 - \xi_3}  
    \frac{ 2 {\xi_2}+{\xi_3}}{12 {\xi_1}}   
    \Big(   f(  {\xi_2}) -  f(  {\xi_3}) \Big)  
\nonumber\\&&\mbox{}  
    + \frac{1}{\xi_2 - \xi_3}  
    \frac{{\xi_2}}{2 {\xi_1}}   
\left(     
    \frac{ f(  {\xi_2}) -1 }{  {\xi_2}} -   
    \frac{ f(  {\xi_3}) -1 }{  {\xi_3}}  
\right),         
\\[3mm]  
M_3 (\xi_1, \xi_2, \xi_3) &= & \mbox{}  
    - F( \xi_1,  \xi_2,  \xi_3)  
    \frac{{\xi_2} {\xi_3}}{\Delta}   
\nonumber\\&&\mbox{}  
    + f(  {\xi_1}) \frac{1}{2 D}(2 {\xi_3} - {\xi_1})  
\nonumber\\&&\mbox{}  
    - f(  {\xi_2}) \frac{1}{2 {\xi_1} D}   
    (2 {\xi_1} {\xi_3} + {\xi_2}^2 - {\xi_3}^2 - {\xi_1}^2)  
\nonumber\\&&\mbox{}  
    + \frac{1}{\xi_2 - \xi_3}   
    \frac{ {\xi_2} }{2 {\xi_1}}   
    \Big(    f(  {\xi_2})  - f(  {\xi_3}) \Big),                   
\\[3mm]    
M_4 (\xi_1, \xi_2, \xi_3) &= &   
    \frac13 F( \xi_1,  \xi_2,  \xi_3)   
    - f(  {\xi_1}) \Big( \frac{1}{{\xi_3}}   
    - \frac{{\xi_1}}{2 {\xi_2}{\xi_3}}  \Big).  
\end{eqnarray}

  
\section*{Appendix B: Third-order form factors for  
computation of the effective action}  
  
  
Here is the list of third-order form  
factors in Eq.~(\ref{totalD}) of section~\ref{result}  
with $D$ defined by Eq.~(\ref{DD}).  
\begin{eqnarray}  
N_1 (s|\Box_1,\Box_2,\Box_3) &=&  
s F(-s\Box_1,-s\Box_2,-s\Box_3)  
{{ {\Box_1} {\Box_2} {\Box_3}}\over{3 {{D}^2}}}  
\nonumber\\&&\mbox{}  
+ f(-s\Box_1)\frac{1}{8 {{D}^2}{\Box_1} {\Box_2}}  
  ({{\Box_1}^4} - 2 {{\Box_1}^3} {\Box_3}+ 2 {\Box_1} {{\Box_3}^3}  
- {{\Box_3}^4}- 2 {{\Box_1}^3} {{\Box_2}}  
\nonumber\\&&\mbox{}  
+3{\Box_2}{{\Box_3}^3}  
- 8 {{\Box_1}^2}  {\Box_2} {\Box_3}  
+8 {\Box_1} {\Box_2} {{\Box_3}^2}  
- 10 {\Box_1} {{\Box_2}^2} {\Box_3} -2 {{\Box_2}^2} {{\Box_3}^2})  
\nonumber\\&&\mbox{}  
-\left(\frac{f(-s\Box_1)-1}{s{\Box_1}}\right)  
\frac{1}{4 D{\Box_1} {\Box_2}}  
( {{\Box_1}^2} + 4 {\Box_1} {\Box_2}  
+ {\Box_2} {\Box_3} - {{\Box_3}^2} )  
\nonumber\\&&\mbox{}  
+ \frac{1}{{\Box_2}-{\Box_3}}\frac{\Box_2}{\Box_1}\left[  
-\frac18\left(\frac1{\Box_2}f(-s\Box_2)  
-\frac1{\Box_3}f(-s\Box_3)\right)  
\right.  
\nonumber\\&&\mbox{}  
\left.  
+\frac14  
\left(\frac1{\Box_2}\frac{f(-s\Box_2)-1}{s{\Box_2}}  
-\frac1{\Box_3}\frac{f(-s\Box_3)-1}{s{\Box_3}}\right)\right],  
\\[3mm]  
N_2 (s|\Box_1,\Box_2,\Box_3) &=&  
- s F(-s\Box_1,-s\Box_2,-s\Box_3)  
\frac{\Box_3}{D}    
\nonumber\\&&\mbox{}  
+ \frac{\Box_1 - \Box_2 - \Box_3}{D \Box_2}   
f(-s {\Box_1})  
+ \frac{\Box_1 \Box_3 - {\Box_1}^2  
+3 \Box_2 \Box_3 + \Box_1 \Box_2}{D \Box_2 \Box_3}   
f(-s {\Box_3})     
\nonumber\\&&\mbox{}  
-\frac{1}{\Box_1 - \Box_3}  
\frac{\Box_1}{\Box_2}   
\Big( \frac{1}{\Box_1} f(-s {\Box_1}) -    
\frac{1}{\Box_3} f(-s {\Box_3}) \Big),  
\\[3mm]  
N_3 (s|\Box_1,\Box_2,\Box_3) &=&  
s  F(-s\Box_1,-s\Box_2,-s\Box_3) \frac{1}{\Box_1} ,  
\\[3mm]  
N_4 (s|\Box_1,\Box_2,\Box_3) &=&  
- s F(-s\Box_1,-s\Box_2,-s\Box_3)  
\frac{D}{ {\Box_1} {\Box_2} {\Box_3}}.  
\end{eqnarray}

\pagebreak  
  

  
\end{document}